\documentclass[a4paper,11pt]{article}
\usepackage{authblk}
\date{October 4, 2023}
\usepackage{amsmath,upgreek,mathabx}
\usepackage{booktabs,multirow,dcolumn,adjustbox,threeparttable,siunitx,float,graphics,breakcites}
\usepackage{rotating} 
\usepackage{pdflscape}
\usepackage{color}
\usepackage[round]{natbib} 
\usepackage{caption}
\usepackage{subcaption}
\usepackage{hyperref}
\usepackage{etoolbox}
\usepackage[english]{babel}
\AtBeginEnvironment{sidewaystable}{}
\newcolumntype{d}[1]{D{.}{.}{#1}}  
\newcommand\mc[1]{\multicolumn{1}{c}{#1}} 
\newcommand\englishkeywordslabel{Keywords:}
\newcommand\englishkeywords[1]{%
  \begin{list}{}{%
    \setlength{\topsep}{2ex}%
    \settowidth{\leftmargin}{\bfseries\englishkeywordslabel~}%
    \setlength{\labelsep}{0pt}%
    \setlength{\labelwidth}{\leftmargin}%
    \setlength{\itemindent}{0pt}%
  } 
  \raggedright\item[\englishkeywordslabel~]#1
  \end{list}
}

\title{A Markov-switching spatio-temporal ARCH model}
\author[1]{Tzung Hsuen Khoo}
\author[1]{Dharini Pathmanathan}
\author[2]{Philipp Otto}
\author[3]{Sophie Dabo-Niang}
\affil[1]{Institute of Mathematical Sciences, Faculty of
Science, Universiti Malaya, 50603 Kuala Lumpur, Malaysia}
\affil[2]{School of Mathematics and Statistics, University
of Glasgow, United Kingdom}
\affil[3]{CNRS, UMR 8524-Laboratoire Paul Painlevé,
INRIA-MODAL, Université Lille, F-59000 Lille,
France}
        
\begin{document}
\maketitle
\begin{abstract}
Stock market indices are volatile by nature and sudden shocks are known to affect volatility patterns. The autoregressive conditional heteroskedasticity (ARCH) and generalized ARCH (GARCH) models neglect structural breaks triggered by sudden shocks that may lead to an overestimation of persistence, causing an upward bias in the estimates. Different regime-switching models which have abrupt regime-switching governed by a Markov chain were developed to model volatility in financial time series data. Volatility modelling was also extended to spatially interconnected time series resulting in spatial variants of ARCH models. This work proposes a Markov regime-switching framework of the spatio-temporal log-ARCH model. The Markov-switching extension of the model, the estimation procedure and the smooth inferences of the regimes are discussed. The Monte-Carlo simulation studies show that the maximum likelihood estimation method for the proposed model has good finite sample properties. The proposed model was applied on 28 stock indices data which were presumably affected by the 2015-2016 Chinese stock market crash. The results showed that that the proposed model is a better fit compared to that of the one-regime counterpart. Furthermore, the smoothed inference of the data indicated the approximate periods where structural breaks occurred. This model can capture structural breaks which simultaneously occur in locations which are spatially autocorrelated.
\end{abstract}

\englishkeywords{Regime-switching, Structural breaks, Volatility, Stock market, Spatial.}

\section{Introduction}\label{Section 1}
Time series data, such as stock market indices, are known to exhibit volatility clusters throughout a long period. Two of the most popular models on volatility modelling are the autoregressive conditional heteroskedasticity (ARCH) model \citep{engle1982autoregressive} and generalized autoregressive conditional heteroskedasticity (GARCH) model \citep{engle1986modelling}. The GARCH($p$,$q$) model assumes that a variable $y_{t}$ is given by 
\begin{equation}\label{generic GARCH part 1}
    y_{t} = \sigma_{t}\varepsilon_{t}, 
\end{equation}
\begin{equation}\label{generic GARCH part 2}
    \begin{aligned}
    &\sigma_{t}^2 = \omega + \sum_{i=1}^q\alpha_{i}y_{t-i}^2 +  \sum_{i=1}^p\beta_{i}\sigma_{t-i}^2, \\  
    &\omega \geq 0,  \\
    &\alpha_{i} \geq 0, i=1,2,...,q, \\
    &\beta_{i} \geq 0, i=1,2,...,p, \\
    \end{aligned}
\end{equation}
where $\sigma_{t}^2$ is the conditional variance, $\{\varepsilon_{t}\}$ is a sequence of independent and identically distributed (i.i.d) random variables with zero mean and unit variance, $\omega$ is a constant, $\{\alpha_{i}\}$ are the ARCH parameters, $\{\beta_{i}\}$ are the GARCH parameters.

When $p=0$, (\ref{generic GARCH part 2}) becomes an ARCH($q$) model. The conditional variance of a financial time series is governed by a GARCH process with a constant unconditional variance. However, international financial markets are susceptible to sudden large shocks such as the 2007-2008 financial crisis and the Chinese stock market crash (2015-2016). Such shocks can trigger abrupt breaks in the unconditional variance in the time series, which correspond to structural breaks in parameters $\{\alpha_{i}\}$ and $\{\beta_{i}\}$ in GARCH models. It is evident that overlooking such structural breaks can result in an upward bias in the estimated persistence of GARCH models, which is measured by $\sum_{i=1}^q\alpha_{i} + \sum_{i=1}^p\beta_{i}$ \citep{mikosch2004nonstationarities,hillebrand2008chapter}.

The occurrence of structural breaks also has significant implications for predicting the volatility of financial time series. It is common to use an expanding window to estimate the parameters of GARCH forecasting models in out-of-sample volatility forecasting. However, this approach may not be appropriate when structural breaks exist in the time series. \cite{west1995predictive} suggested that the volatility forecasting accuracy of GARCH(1,1) models for exchange rate returns could be achieved by accounting for the occurrence of structural breaks in the unconditional variance of exchange rate returns. \cite{starica2005does} indicated that forecasts of stock return volatility with GARCH(1,1) models assuming stable parameters are often inferior to forecasts by models which can accommodate frequent changes in the unconditional variance of stock returns.

The structural breaks in the parameters of GARCH models can be accounted for by making the parameters in the models to be time-varying. The parameters $\{\alpha_{i}\}$ and $\{\beta_{i}\}$ can be different depending on an unobserved random variable $s_{t}$ \citep{hamilton1994time}. The random variable $s_{t}$ is often denoted as the \textit{state} or \textit{regime}, and this will be referred to \textit{regime} for the remainder of this work. Different regime-switching models, which incorporate abrupt regime-switching governed by a Markov chain, have been developed to model the volatility of financial data from a temporal perspective \citep{hamilton1994autoregressive,gray1996modeling, haas2004new}.

Volatility modelling was also extended to spatially interconnected time series. The spatial ARCH (spARCH) model was introduced by \cite{otto2018generalised}, where its properties were further analysed \citep{otto2021stochastic}. A logarithmic expression of the volatility equation was suggested by \cite{sato2017spatial}. This can be seen as a spatially equivalent log-ARCH model. The spARCH model was then generalized to allow for different spatial GARCH-type models \citep{otto2022general}. Recently, a dynamic spatio-temporal log-ARCH model was proposed, and its estimation procedure and asymptotic properties were provided \citep{otto2022dynamic}. 

The phenomenon of regime-switching can be extended in a spatial and temporal framework where each regime-switching can occur simultaneously across different spatial locations, reflecting the interconnected behaviour of the data. This enables the model to capture intricate dynamics that could be missed by traditional Markov-switching ARCH/GARCH models, where sudden changes in volatility may be both spatially and temporally dependent. The literature on spatially extended Markov-switching model is very limited. To the best of our knowledge, one study by \cite{kondo2022spatial} proposed a two-regime Markov-switching model with a spatial autoregressive process, which was applied to investigate the spatial dependence of regional business cycles in Mexico. However, this model does not account for the ARCH and spatio-temporal effects which will be considered in our model. In this article, we introduce a novel Markov-switching model with a dynamic spatio-temporal log-ARCH process. 

The rest of the paper is structured as follows. Section \ref{Section 2} discusses the dynamic spatio-temporal log-ARCH model \citep{otto2022dynamic}, the Markov-switching extension of the model, the estimation procedure, and the smooth inference of regimes. In section \ref{Section 3}, a Monte Carlo simulation study is presented to investigate the finite sample properties of the maximum likelihood estimation. In section \ref{Section 4}, applications on global stock market data are investigated. Section \ref{Section 5} concludes with recommendations for future studies. 

\section{Methodology}\label{Section 2}
\subsection{Dynamic spatio-temporal ARCH Model}\label{Section 2.1}
In this work, we propose a regime-switching version of the spatio-temporal ARCH models developed by \cite{otto2022dynamic}. Thus, we briefly outline the single-regime spatio-temporal ARCH model with one spatial lag as below:
\begin{equation}
  \label{starch_part1}
   y_{it} = h_{it}^\frac{1}{2}\varepsilon_{it},
\end{equation}
\begin{equation}
  \label{starch_part2}
   \log{h_{it}} = \sum_{j=1}^{n}{\rho_{0}w_{ij}\log{y_{jt}^2}} + 
                  \gamma_{0}\log{y_{it-1}^2} + 
                  \sum_{j=1}^{n}{\delta_{0}w_{ij}\log{y_{jt-1}^2}} + \mu_{0},
\end{equation}
for $i= 1,2,...,n$ and $t= 1,2,...,T$. \\
In this model, $y_{it}$ is a variable of region $i$ and time $t$, $h_{it}$ is defined as the volatility term, $w_{ij}$ is the $ij$-th entry of the $n$ x $n$ spatial weight matrix $W_{n}$, and $\{\varepsilon_{it}\}$ are i.i.d. random variables across $i$ and $t$ with a mean of zero and unit variance; $\mu_{0}$ is the regional fixed effect. Lastly, $(\rho_{0}, \gamma_{0}, \delta_{0})$ are the measures of spatial, temporal, and spatio-temporal effects, respectively.

In contrast to the original model, we do not consider any exogenous variables or temporal fixed effects. Moreover, the regional fixed effect for each location is assumed to be the same. A log-square operation on both sides of (\ref{starch_part1}) and arranging $n$ variables in a vector form leads to:

\begin{equation}\label{vector_starch_part1}
  \mathbf{Y}_{t}^* = \mathbf{h}_{t}^* + \boldsymbol{\mathcal{E}}_{t}^*,
\end{equation}
\begin{equation}\label{vector_starch_part2}
     \mathbf{h}_{t}^* = \rho_{0}W_{n}\mathbf{Y}_{t}^* +
                     \gamma_{0}\mathbf{Y}_{t-1}^* +  
                     \delta_{0}W_{n}\mathbf{Y}_{t-1}^* + \mu_{0}\mathbf{1}_{n} + 
                     \boldsymbol{\mathcal{E}}_{t}^*, 
\end{equation}
where 
$\mathbf{Y}_{t}^*$ = $(\log{y_{1t}^2},\log{y_{2t}^2},...,\log{y_{nt}^2})'$, 
$\mathbf{h}_{t}^*$ = $(\log{h_{1t}},\log{h_{2t}},...,\log{h_{nt}})'$, 
$\boldsymbol{\mathcal{E}}_{t}^*$ = $(\log{\varepsilon_{1t}^2},\log{\varepsilon_{2t}^2},...,\log{\varepsilon_{nt}^2})'$ and $\mathbf{1}_{n}$ is a $n$ x $1$ vector of ones. Substituting (\ref{vector_starch_part2}) into (\ref{vector_starch_part1}) results in
\begin{equation}\label{full vector starch}
    \mathbf{Y}_{t}^* = \rho_{0}W_{n}\mathbf{Y}_{t}^* + \gamma_{0}\mathbf{Y}_{t-1}^* +  \delta_{0}W_{n}\mathbf{Y}_{t-1}^* + \mu_{0}\mathbf{1}_{n} + \boldsymbol{\mathcal{E}}_{t}^*.
\end{equation}
The entries of $\boldsymbol{\mathcal{E}}_{t}^*$ in (\ref{full vector starch}) are still i.i.d across $i$ and $t$ but the means may not be zero. Thus, $E[log{ \varepsilon_{i,t}^2}]$ is added and subtracted in (\ref{full vector starch}):
\begin{align}\label{full vector starch, 2}
    \mathbf{Y}_{t}^* &= \rho_{0}W_{n}\mathbf{Y}_{t}^* + \gamma_{0}\mathbf{Y}_{t-1}^* +  \delta_{0}W_{n}\mathbf{Y}_{t-1}^* + \mu_{0}\mathbf{1}_{n} + \mu_{\varepsilon}\mathbf{1}_{n} + \boldsymbol{\mathcal{U}}_{t}^* \nonumber\\
    & = \rho_{0}W_{n}\mathbf{Y}_{t}^* + \gamma_{0}\mathbf{Y}_{t-1}^* +  \delta_{0}W_{n}\mathbf{Y}_{t-1}^* + \phi_{0}\mathbf{1}_{n} + \boldsymbol{\mathcal{U}}_{t}^*
\end{align}
where  $\phi_{0}=\mu_{0} + \mu_{\varepsilon},\, \mu_{\varepsilon} = E[log{\varepsilon_{i,t}^2}],\,\boldsymbol{\mathcal{U}}_{t}^* =(\log{\varepsilon_{1t}^2} - E[log{\varepsilon_{i,t}^2}],\,\log{\varepsilon_{2t}^2} 
 - E[log{\varepsilon_{i,t}^2}],\,...\,,\,\log{\varepsilon_{nt}^2}-E[log{\varepsilon_{i,t}^2}])'$.

\subsection{Markov-switching spatio-temporal ARCH model}\label{Section 2.2}
In this section, we describe the regime-switching version of (\ref{full vector starch, 2}). Let the state variable $s_{t}$ be a random variable which only takes on integers $\{1,2,...,K\}$. If the conditional probability $P\{s_{t} = i_{t}|s_{t-1} = i_{t-1}, s_{t-2} = i_{t-2},...\}$ equals to $P\{s_{t} = i_{t}|s_{t-1} = i_{t-1}\}$, then $s_{t}$ can be defined as a $K$-state Markov chain with a $K$ x $K$ transition probability matrix $\mathbf{P}_{n}$ ~\citep{hamilton1994time}:
\[
  \mathbf{P}_{n} = \begin{pmatrix}
    p_{11} & p_{21} & \cdots & p_{K1} \\
    p_{21} & p_{22} & \cdots & p_{K2} \\
    \vdots & \vdots & \vdots & \vdots \\
    p_{K1} & p_{K2} & \cdots & p_{KK}
  \end{pmatrix},
\]
where $p_{ij}$ is the transition probability from state $i$ to state $j$. $\{p_{ij}\}_{i,j=1,...,K}$ also satisfies the following relation:
\begin{equation}\label{sum_equals_1}
 p_{i1} + p_{i2} + \cdots + p_{iK} = 1.
\end{equation}
In this work, a two-regime Markov-switching spatio-temporal ARCH model will be introduced. A two-regime Markov-switching spatio-temporal ARCH model can be formulated as follows:
\begin{equation}\label{2Regime STARCH}
    \mathbf{Y}_{t}^* = \rho_{0,s_{t}}W_{n}\mathbf{Y}_{t}^* + \gamma_{0,s_{t}}\mathbf{Y}_{t-1}^* +  \delta_{0,s_{t}}W_{n}\mathbf{Y}_{t-1}^* + \phi_{0}\mathbf{1}_{n} + \boldsymbol{\mathcal{U}}_{t}^*,
\end{equation}
where the state variable $s_{t}$ follows a 2-state Markov chain with $2$ x $2$ transition probability matrix $P$,
\begin{equation}\label{2x2 Transition Matrix}
\mathbf{P} = \begin{pmatrix}
    p & 1-q  \\
    1-p & q 
  \end{pmatrix},
\end{equation}
and therefore, assumes the value of $\{1,2\}$. $s_{t}$ = 1 and $s_{t}$ = 2 are equivalent to regime 1 and 2 respectively. Even though the distribution of $\boldsymbol{\mathcal{U}}_{t}^*$ is skewed, we assume $\boldsymbol{\mathcal{U}}_{t}^*$ to be $N(\mathbf{0}_{n}, \sigma_{0}^2\mathbf{I}_{n})$ to derive the quasi-maximum likelihood estimator (QMLE) in the following section.

\subsection{Quasi Maximum likelihood estimation (QMLE)}\label{Section 2.3}
QMLE requires the log-likelihood of $\boldsymbol{\theta}$ = $(\theta_{s_{t}=1},\,\theta_{s_{t}=2},\,p,\,q)'$, where $\theta_{s_{t}=j}$ = $(\rho_{0,s_{t}},\,\gamma_{0,s_{t}},\,\delta_{0,s_{t}},\,\mu_{0,s_{t}})_{j=1,2}$, to be calculated. To this end, the approach of the Hamilton filter~\cite{hamilton1989new} is followed. Firstly, the $2$ x $1$ vector $\boldsymbol{\eta}_{t}$ contains the conditional densities of $\mathbf{Y}_{t}^*$ is defined as follows:
\begin{align}\label{cond.densities}
  \boldsymbol{\eta}_{t} &= \begin{pmatrix}
    f(\mathbf{Y}_{t}^*|\, s_{t} = 1, \mathbf{Y}_{t-1}^*;\boldsymbol{\theta}) \\[0.5ex]
    f(\mathbf{Y}_{t}^*|\, s_{t} = 2, \mathbf{Y}_{t-1}^*;\boldsymbol{\theta})
  \end{pmatrix}.
\end{align}
The conditional densities of $\mathbf{Y}_{t}^*$ can be obtained by first rearranging (\ref{full vector starch}),
\begin{align}
    \boldsymbol{\mathcal{U}_{t}^*} = (\mathbf{I}_{n}-\rho_{0,s_{t}}W_{n})\mathbf{Y}_{t}^* - \gamma_{0,s_{t}}\mathbf{Y}_{t-1}^* - \delta_{0,s_{t}}W_{n}\mathbf{Y}_{t-1}^* - \phi_{0}\mathbf{1}_{n}.
    \nonumber
\end{align}
Next, $\boldsymbol{\mathcal{U}_{t}^*}$ is transformed into $\mathbf{Y}_{t}^*$:
\begin{align}
     f(\mathbf{Y}_{t}^*|\,s_{t}, \mathbf{Y}_{t-1}^*;\boldsymbol{\theta}) &= f(\boldsymbol{\mathcal{U}_{t}^*})\Big\lvert\frac{\partial\boldsymbol{\mathcal{U}_{t}^*}}{\partial\mathbf{Y}_{t}^*}\Big\rvert
    \\ & = \dfrac{1}{(2\pi\sigma_{0}^2)^{-n/2}}\exp\Big[-\dfrac{\boldsymbol{\mathcal{U}_{t}^*}'\boldsymbol{\mathcal{U}_{t}^*}}{2\sigma_{0}^2}\Big]\big\lvert \mathbf{I}_{n}-\rho_{0,s_{t}}W_{n}\big\rvert.
     \nonumber
\end{align}
(\ref{cond.densities}) can then be expressed as follows,
\begin{align}\label{cond.densities, full}
\boldsymbol{\eta}_{t} &= \begin{pmatrix}
     \dfrac{\big\lvert \mathbf{I}_{n} -\rho_{0,1}W_{n}\big\rvert}{(2\pi\sigma_{0}^2)^{-n/2}}\exp\Big[-\dfrac{\boldsymbol{\mathcal{U}_{t}^*}'\boldsymbol{\mathcal{U}_{t}^*}}{2\sigma_{0}^2}\Big] \\[2ex]
     \dfrac{\big\lvert \mathbf{I}_{n} -\rho_{0,2}W_{n}\big\rvert}{(2\pi\sigma_{0}^2)^{-n/2}}\exp\Big[-\dfrac{\boldsymbol{\mathcal{U}_{t}^*}'\boldsymbol{\mathcal{U}_{t}^*}}{2\sigma_{0}^2}\Big] 
  \end{pmatrix}.
\end{align}
Secondly, we are required to also construct the $2$ x $1$ vectors of filtered probability $\boldsymbol{\upxi}_{t|t}$, and predicted probability $\boldsymbol{\xi}_{t+1|t}$:
\begin{equation}\label{filt.probabilities}
  \boldsymbol{\xi}_{t|t} = \begin{pmatrix}
    P\{s_{t} = 1|\mathbf{Y}_{t}^*;\boldsymbol{\theta}\} \\ 
    P\{s_{t} = 2|\mathbf{Y}_{t}^*;\boldsymbol{\theta}\} 
  \end{pmatrix},
\end{equation}
\begin{equation}\label{pred.probabilities}
  \boldsymbol{\xi}_{t+1|t} = \begin{pmatrix}
    P\{s_{t+1} = 1|\mathbf{Y}_{t}^*;\boldsymbol{\theta}\} \\ 
    P\{s_{t+1} = 2|\mathbf{Y}_{t}^*;\boldsymbol{\theta}\} 
  \end{pmatrix}.
\end{equation}
By assuming that $\boldsymbol{\theta}$ is known, the idea of the Hamilton filter is to generate optimal inference and prediction for each $t$ by iterating the following equations,
\begin{equation}\label{Hamiltonfilter.part1}
   \boldsymbol{\xi}_{t|t} = \dfrac{\boldsymbol{\xi}_{t|t-1}\bigodot\boldsymbol{\eta}_{t}}{\boldsymbol{\xi}_{t|t-1}\cdot\boldsymbol{\eta}_{t}},
\end{equation}
\begin{equation}\label{Hamiltonfilter.part2}
   \boldsymbol{\xi}_{t+1|t} = \mathbf{P}\cdot\boldsymbol{\xi}_{t|t},
\end{equation}
where $\mathbf{P}$ represents the $2$ x $2$ transition probability matrix defined in (\ref{2x2 Transition Matrix}), $\bigodot$ denotes element-by-element multiplication. Once the iterations are completed for $t=1,2,...,T$, the log-likelihood function $\mathcal{L(\boldsymbol{\theta})}$ can be calculated as:
\begin{equation}\label{Log-likelihood equation}
  \mathcal{L(\boldsymbol{\theta})} = \sum_{t=1}^{T-1}\log{f(\mathbf{Y}_{t}^*|\,s_{t}, \mathbf{Y}_{t-1}^*;\boldsymbol{\theta})},
\end{equation}
where $\log{f(\mathbf{Y}_{t}^*|\,s_{t}, \mathbf{Y}_{t-1}^*;\boldsymbol{\theta})}$ is expressed by 
\begin{equation}\label{last step of filt.probs}
    \boldsymbol{\xi}_{t+1|t} = \mathbf{1}'(\boldsymbol{\xi}_{t|t-1}\bigodot\boldsymbol{\eta}_{t}).
\end{equation}
The log-likelihood in (\ref{Log-likelihood equation}) is then maximized numerically using the R package \texttt{bbmle}. ~\citep{bolker2017package}.
\subsubsection{Constraints on Parameters in QMLE}
Constraints on the parameters are necessary to ensure the numerical maximization is well-behaved. In general, $\rho_{0,s_{t}} \in$ ($\lambda_{min}^{-1}$,$\lambda_{max}^{-1}$), where $\lambda$ is an eigenvalue of $W_{n}$, leads to a non-singular $I_{n}-\rho_{0,s_{t}}W_{n}$ when $W_{n}$ is symmetric, and a positive definite variance-covariance matrix \cite{ord1975estimation}. When $W_{n}$ is row-normalized, the parameter can be further simplified to (-1, 1). Secondly, to ensure weak stationarity across time, we assume that $ -1< \gamma_{0,s_{t}} < 1$. Furthermore, $ -1 < \delta_{0,s_{t}} < 1$ and the sum of $\rho_{0,s_{t}}$ and $\delta_{0,s_{t}}$ needs to smaller than one for each regime \citep[see][for more details on the parameter space]{otto2022dynamic}.
We employ transformations to constrain the parameters\citep{kim1999state}. For example, the transformation of a parameter $x$ is expressed by 
\begin{equation}
    x = a+\exp{x}', \nonumber 
\end{equation}
when $x>a$. On the other hand, if $a<x<b$, then the transformation of $x$ can be done by 
\begin{equation}
    x = a+ \frac{(b-a)}{1+\exp(-x')}. \nonumber 
\end{equation}

\subsection{Smoothed inference}\label{Section 2.4}
A form of inference on the state variable $s_{t}$ of the vector $\mathbf{Y}_{t}^*$ was introduced in section~\ref{Section 2.2} in the form of filtered probability (\ref{filt.probabilities}). The iterative calculation of $\boldsymbol{\xi}_{t|t}$ from (\ref{cond.densities}) to (\ref{last step of filt.probs}) uses observation vectors $\mathbf{Y_{\tau}^*}, \tau=1,...,t$. Another form of inference on state variable $s_{t}$, which is defined as the smoothed inference, uses data obtained at time t < $\tau$. Smoothed inference will be used in section \ref{Section 4} to infer the state variables $s_{t}, t=1,2,...,T$ of the data set. Smoothed inferences are calculated using an iterative algorithm by \cite{kim1994dynamic}, and it is formulated as:
\begin{equation}\label{smooth probability equation}
   \boldsymbol{\xi}_{t|T} = \boldsymbol{\xi}_{t|t} \bigodot\{\mathbf{P'}\cdot[\boldsymbol{\xi}_{t+1|T} \bigodiv \boldsymbol{\xi}_{t+1|t} ]\},
\end{equation}
where $\bigodiv$ denotes the operation of element-by-element division. In a two-regime framework, the $2$ x $1$ smoothed probabilities $\boldsymbol{\xi}_{t|T}$ can then be obtained by a backward iteration for $t=T-1,T-2,..,1$ using (\ref{smooth probability equation}). 

\section{Monte Carlo simulation studies}\label{Section 3}
\begin{figure}[!ht]
     \centering
     \begin{subfigure}[b]{0.55\textwidth}
         \centering
         \includegraphics[width=\textwidth]{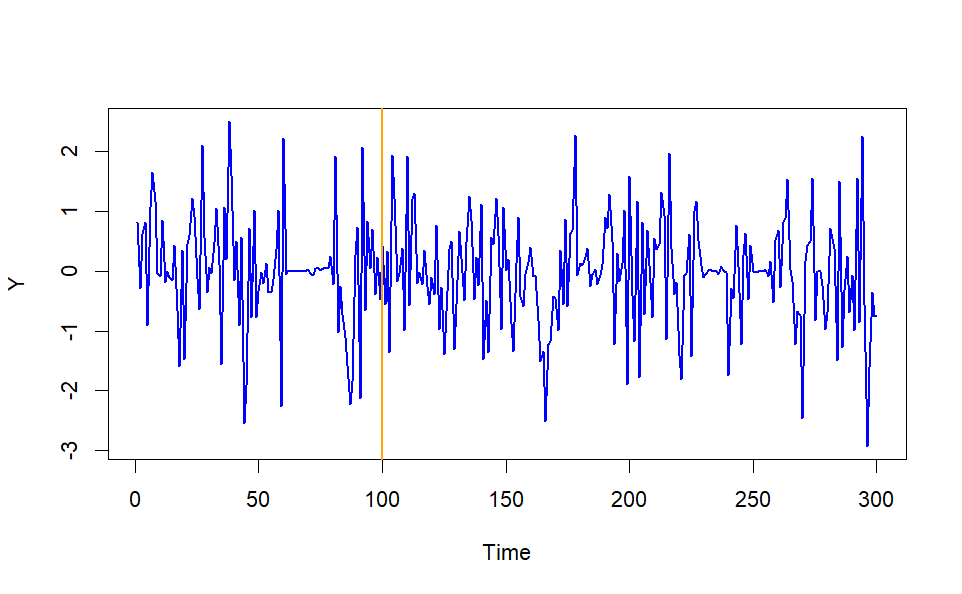}
         \label{fig (a)}
         \caption{}
     \end{subfigure}
     \begin{subfigure}[b]{0.44\textwidth}
         \centering
         \includegraphics[width=\textwidth]{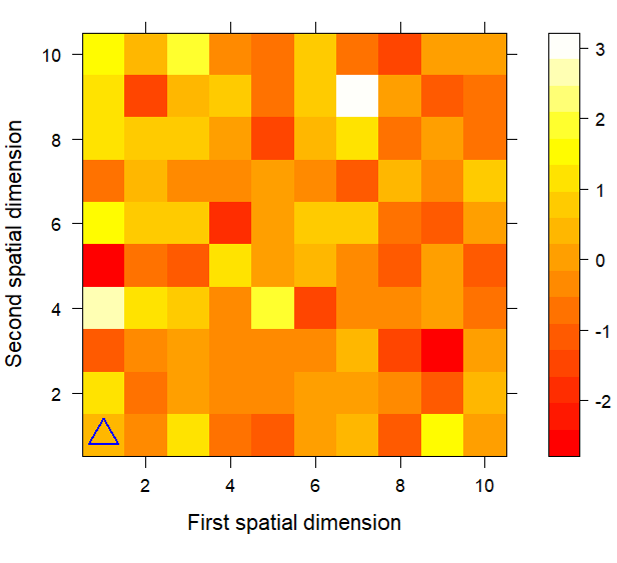}
         \label{fig (b)}
         \caption{}
     \end{subfigure}
     \caption{(a) An example of a simulated time series for the location $(1,1)$ in a $10$ x $10$ grid with $T$ = 300 (b) A plot of simulated values in a $10$ x $10$ grid at the time point = 100; $\bigtriangleup$ corresponds to the value indicated by the vertical line in (a).}
     \label{fig:example simulated time series}
\end{figure}

We performed Monte Carlo simulation studies to investigate the finite sample properties of the QMLE of the two-regime Markov-switching spatio-temporal ARCH model discussed in section \ref{Section 2.3}. Data required in this simulation study are generated with the error terms $\{\varepsilon_{i,t}\}, i = 1,...,n, t = 1,...,T$ randomly generated from $N.I.I.D(0,1)$, and a row-normalized spatial weight matrix $W_{n}$ generated using Queen contiguity. The parameters $(\rho_{0,1},\,\gamma_{0,1},\,\delta_{0,1},\,\mu_{0,1},\,p)$ and $(\rho_{0,2},\,\gamma_{0,2},\,\delta_{0,2},\,\mu_{0,2},\,q)$ are set to be (0.2,\,0.2,\,-0.2,\,0.1,\,0.1,\,0.97) and (0.2,\,0.8,\,-0.2,\,0.1,\,0.93) respectively. The parameters are chosen to simulate a weakly temporal-dependent model and a persistent model, respectively. Using the above-mentioned parameters, two plots are generated and illustrated in Figure \ref{fig:example simulated time series}. The first plot is an example of the simulated time series at location (1,1) in a $10 \times 10$ grid; the second plot is a plot of simulated values of 100 locations at time point = 100. 

We performed Monte Carlo simulation studies based on the sample size $n$ and time period $T$ of $\{(n,T) = (36,200),(36,300),(36,500),(49,200),(49,300),\\(49,500),(100,200),(100,300),(100,500)\}$ with 100 repetitions. The empirical means and root-mean-square errors (RMSE) for the estimators are shown in Table~\ref{table:1}. The results of the simulation studies show that the empirical means and RMSE improve as the sample size and time period $(n,T)$ increase from $(36,200)$ to $(100,500)$. This suggests the mean converges to the true value as both $n$ and $T$ increase. In summary, the suggested estimation procedure exhibits good finite sample properties.

\begin{sidewaystable}
\captionsetup{singlelinecheck = false, format= hang, justification=raggedright}
\caption{The empirical means and RMSE (bracketed) of the estimators in the Monte Carlo simulation studies.}\label{table:1}
\setlength\tabcolsep{3pt}
\begin{tabular}{@{\extracolsep\fill} l d{2.3} d{2.3} d{2.3} d{2.3} d{2.3} d{2.3} d{2.3} d{2.3} d{2.3} @{\extracolsep\fill}}
\toprule
  \multirow{2}{*}{\textbf{Parameter}}  & \multicolumn{3}{c}{${n=36}$} & \multicolumn{3}{c}{${n=49}$}
   & \multicolumn{3}{c}{$n=100$} \\
  \cmidrule(lr){2-4} \cmidrule(lr){5-7} \cmidrule(lr){8-10}
   & \mc{\parbox{2cm}{\centering {$T=200$}}} & \mc{\parbox{2cm}{\centering {$T=300$}}} & \mc{\parbox{2cm}{\centering {$T=500$}}}
   & \mc{\parbox{2cm}{\centering {$T=200$}}} & \mc{\parbox{2cm}{\centering {$T=300$}}} & \mc{\parbox{2cm}{\centering {$T=500$}}}
   & \mc{\parbox{2cm}{\centering {$T=200$}}} & \mc{\parbox{2cm}{\centering {$T=300$}}} & \mc{\parbox{2cm}{\centering {$T=500$}}} \\
  \midrule 
$\rho_{0,1}=0.2$    
&\mc{0.199(0.028)} &\mc{0.207(0.075)} &\mc{0.199(0.016)}    
&\mc{0.197(0.020)} &\mc{0.197(0.014)} &\mc{0.201(0.012)}   
&\mc{0.197(0.014)} &\mc{0.199(0.010)} &\mc{0.200(0.009)}   
\\
$\gamma_{0,1}=0.2$  
&\mc{0.204(0.046)} &\mc{0.203(0.048)} &\mc{0.198(0.009)}    
&\mc{0.197(0.012)} &\mc{0.198(0.009)} &\mc{0.200(0.007)}   
&\mc{0.200(0.007)} &\mc{0.199(0.006)} &\mc{0.200(0.005)}
\\ 
$\delta_{0,1}=-0.2$  
&\mc{-0.193(0.037)} &\mc{-0.204(0.031)} &\mc{-0.199(0.023)}  
&\mc{-0.196(0.024)} &\mc{-0.199(0.020)} &\mc{-0.200(0.014)} 
&\mc{-0.200(0.016)} &\mc{-0.199(0.013)} &\mc{-0.200(0.011)}   
\\    
$\mu_{0,1}=0.1$     
&\mc{0.116(0.123)}  &\mc{0.095(0.058)} &\mc{0.100(0.038)}   
&\mc{0.105(0.052)}  &\mc{0.094(0.040)} &\mc{0.102(0.031)}   
&\mc{0.096(0.035)}  &\mc{0.100(0.031)} &\mc{0.099(0.025)}   
\\   
$p=0.97$            
&\mc{0.968(0.016)} &\mc{0.966(0.13)}  &\mc{0.967(0.009)}    
&\mc{0.966(0.016)} &\mc{0.967(0.011)} &\mc{0.967(0.010)}   
&\mc{0.970(0.015)} &\mc{0.969(0.014)} &\mc{0.968(0.009)}
\\  
$\rho_{0,2}=0.2$    
&\mc{0.204(0.079)} &\mc{0.198(0.034)} &\mc{0.206(0.035)}    
&\mc{0.193(0.065)} &\mc{0.202(0.030)} &\mc{0.199(0.022)}   
&\mc{0.206(0.027)} &\mc{0.201(0.025)} &\mc{0.200(0.018)}   
\\  
$\gamma_{0,2}=0.8$  
&\mc{0.797(0.027)} &\mc{0.795(0.027)} &\mc{0.800(0.009)}    
&\mc{0.800(0.013)} &\mc{0.801(0.011)} &\mc{0.799(0.006)}   
&\mc{0.799(0.010)} &\mc{0.801(0.007)} &\mc{0.799(0.004)} 
\\
$\delta_{0,2}=-0.2$  
&\mc{-0.193(0.126)} &\mc{-0.196(0.039)} &\mc{-0.207(0.042)}  
&\mc{-0.197(0.068)} &\mc{-0.204(0.034)} &\mc{-0.199(0.025)}   
&\mc{-0.211(0.036)} &\mc{-0.203(0.030)} &\mc{-0.201(0.021)} 
\\    
$\mu_{0,2}=0.1$     
&\mc{0.093(0.326)} &\mc{0.093(0.139)} &\mc{0.098(0.095)}    
&\mc{0.077(0.142)} &\mc{0.094(0.101)} &\mc{0.095(0.062)}   
&\mc{0.075(0.088)} &\mc{0.097(0.076)} &\mc{0.091(0.048)} 
\\    
$q=0.93$             
&\mc{0.893(0.087)} &\mc{0.913(0.039)} &\mc{0.919(0.032)}   
&\mc{0.903(0.056)} &\mc{0.908(0.046)} &\mc{0.922(0.029)}   
&\mc{0.914(0.049)} &\mc{0.918(0.033)} &\mc{0.927(0.024)} 
\\ 
\bottomrule
\end{tabular}
\end{sidewaystable}

\section{Application to real data}\label{Section 4}
We apply the method to 26 Asian stock indices and two United States (US) stock indices during the period from 1 January 2014 to 1 January 2019. During this period, there was a global stock market sell-off that spanned from around June 2015 to June 2016, triggered by the sharp decline of the Shanghai Stock Exchange in China. This event was subsequently compounded by other significant global events, including the Greek debt default in June 2015, the decrease in oil prices, and the revelation of the Brexit vote in February 2016. The effect of the Chinese stock market sell-off (2015-2016) on the spatial autocorrelation of 71 global stock indices was investigated by \cite{khoo2023spatial}, and the results suggested that the occurrence of the stock market sell-off increased spatial autocorrelations among the stock indices. The present work applies the newly developed Markov-switching spatio-temporal ARCH model to investigate a smaller subset of stock indices, comprising most of the Asian stock indices, along with two US stock indices within a spatio-temporal framework. 

The main descriptive statistics of the selected stock indices are illustrated in Table \ref{table:2}. The daily closing prices are acquired from \url{http://www.investing.com}, and the log-returns of the daily closing prices are calculated before the application of the model. The proposed application to real data seeks to investigate the occurrence of structural breaks due to the formation of two regimes during the time period of interest.  

\begin{center}
\begin{table}[!t]%
\captionsetup{singlelinecheck = false}
\caption{The main descriptive statistics of 28 stock indices in the selected time period.}
\label{table:2}
\setlength\tabcolsep{0pt}
\begin{tabular*}{\textwidth}{@{\extracolsep\fill}  ll *4{d{2.5}} @{\extracolsep\fill}}
\toprule
\textbf{Stock Index}       & \textbf{Country}          & \mc{\textbf{Mean}}     &\mc{\textbf{St. Dev.}} &\mc{\textbf{Min}}  &\mc{\textbf{Max}} \\ \midrule
NASDAQ            & USA              & 0.00048  & 0.00892  & -0.04216 & 0.04152 \\
NYSE Composite    & USA              & 0.00022  & 0.00753  & -0.04396 & 0.02921 \\
SCI               & China            & 0.00044  & 0.01511  & -0.08873 & 0.05604 \\
SZ Component      & China            & 0.00031  & 0.01740  & -0.08603 & 0.06254 \\
HSI               & Hong Kong        & 0.00026  & 0.01025  & -0.06018 & 0.04021 \\
TAIEX             & Taiwan           & 0.00021  & 0.00762  & -0.04957 & 0.03518 \\
KOSPI             & South Korea      & 0.00022  & 0.00681  & -0.03143 & 0.02912 \\
NIKKEI 225        & Japan            & 0.00032  & 0.01272  & -0.08253 & 0.07426 \\
JKSE              & Indonesia        & 0.00037  & 0.00833  & -0.04088 & 0.04451 \\
KLSE              & Malaysia         & -0.00004 & 0.00525  & -0.02738 & 0.02222 \\
STI               & Singapore        & 0.00007  & 0.00697  & -0.04390 & 0.02656 \\
SET               & Thailand         & 0.00034  & 0.00731  & -0.04842 & 0.04484 \\
PSEI              & Philippine       & 0.00034  & 0.00889  & -0.06939 & 0.03576 \\
VN-30             & Vietnam          & 0.00055  & 0.00929  & -0.05775 & 0.04144 \\
XAO               & Australia        & 0.00013  & 0.00790  & -0.04115 & 0.03204 \\
NZX-50            & New Zealand      & 0.00055  & 0.00513  & -0.03395 & 0.02412 \\
SENSEX-30         & India            & 0.00046  & 0.00824  & -0.06120 & 0.03324 \\
NIFTY-50          & India            & 0.00050  & 0.00832  & -0.06097 & 0.03312 \\
CSE               & Sri Lanka        & 0.00007  & 0.00472  & -0.02699 & 0.01758 \\
KARACHI-100       & Pakistan         & 0.00045  & 0.00903  & -0.04765 & 0.04419 \\
ASE               & Jordan           & 0.00002  & 0.00397  & -0.01981 & 0.02086 \\
BSE               & Bahrain          & 0.00006  & 0.00450  & -0.02842 & 0.02754 \\
QE                & Qatar            & -0.00019 & 0.01126  & -0.07543 & 0.07310 \\
ADX               & UAE              & 0.00002  & 0.01029  & -0.07155 & 0.06487 \\
TASI              & Saudi Arabia     & -0.00018 & 0.01208  & -0.07547 & 0.08547 \\
MSM-30            & Oman             & -0.00029 & 0.00676  & -0.06413 & 0.05370 \\
XU-100            & Turkey           & 0.00054  & 0.01238  & -0.07347 & 0.05256 \\
IMOEX             & Russia           & 0.00041  & 0.04183  & -0.25228 & 0.20400 \\ 
\bottomrule
\end{tabular*}
\end{table}
\end{center}

\subsection{Construction of spatial weight matrices}
The spatial weight matrix in the model requires the distance between each stock log-return $Y_t(s_{i}), i=1,2,...,n$ to be defined. Since stock returns do not have a meaningful geographical location, a distance metric based on geographical distances may not be suitable for constructing a spatial weight matrix in the Markov-switching spatio-temporal ARCH  model. Instead, the distance within the financial network could be considered (see, e.g., \citealt{fulle2023spatial,mattera2023network}). For this reason, a model-based approach with individual uni-variate log-ARCH models is adopted in the application. In a model-based approach, each time series is assumed to follow a parametric model such as the ARIMA models. A log-ARCH($p$) model for each log-return $Y_t(s_{i})$ can be formulated as follows:
\begin{equation}\label{uni-logARCH part 1}
    Y_t(s_{i}) = \sqrt{h_t(s_{i})}\varepsilon_{t}(s_{i}),
\end{equation}
\begin{equation}\label{uni-logARCH part 2}
    \log{h_t(s_{i})} = \omega_{i} + \sum_{p=1}^P\gamma_{ip}\log{Y_{t-p}^2(s_{i})},
\end{equation}
where $p$ is the order of the log-ARCH($p$) model, $\omega_{i}$ is the constant term, and $\gamma_{i1},\gamma_{i2},...,\gamma_{ip}$ are the ARCH parameters for $Y_{t}(s_{i})$. A log-square transformation on (\ref{uni-logARCH part 1}), it follows that, 
\begin{equation}\label{log square uni-logARCH part 1}
    \log{Y_t^2(s_{i})} = \omega_{i} + \sum_{p=1}^P\gamma_{ip}\log{Y_{t-p}^2(s_{i})} + \log{\varepsilon_{t}^2(s_{i})},
\end{equation}
\begin{equation}\label{log square uni-logARCH part 2}
    \log{Y_t^2(s_{i})} = \varphi_{i} + \sum_{p=1}^P\gamma_{ip}\log{Y_{t-p}^2(s_{i})} + u_{t}^2(s_{i}),
\end{equation}
where the transformed constant term $\varphi_{i}=\omega_{i} + E(\log{\varepsilon_t^2(s_{i})})$, and $\gamma_{i1},\gamma_{i2},...,\gamma_{ip}$ can now be viewed as autoregressive (AR) parameters for log-squared $Y_t(s_{i})$. The transformation from (\ref{uni-logARCH part 1}) and (\ref{uni-logARCH part 2}) to (\ref{log square uni-logARCH part 1}) and (\ref{log square uni-logARCH part 2}) ensures that $u_t^2(s_i)$ has a mean of zero which allows for consistent estimations of the ARCH parameters.  Once each log-ARCH($p$) model is fitted, the estimated parameters $\gamma_{i1},\gamma_{i2},...,\gamma_{ip}$ are used to measure the dissimilarity for each pair of time series $(Y_{t}(s_{i}),Y_{t}(s_{j}))$. In the application, the Piccolo distance metric \citep{piccolo1990distance}, which is defined as follows, is selected as the dissimilarity measure:
\begin{equation}\label{Piccolo Distance}
    d_{PIC}(Y_{t}(s_{i}),Y_{t}(s_{j})) =\sqrt{\sum_{p=1}^P(\gamma_{ip}-\gamma_{jp})^2}.
\end{equation}
The Piccolo distance for each pair of $(Y_{t}(s_{i}),Y_{t}(s_{j}))$ forms an $n$ x $n$ matrix which is then used to construct the $n$ x $n$ spatial weight matrix $W_{n}$. Each entry $w_{ij}$ of $W_{n}$ is constructed using the criterion of k-nearest neighbours as follows: \\
\begin{equation}
    w_{ij} = \begin{cases}
        \dfrac{1}{\#N_{k}(i)}, & if\hspace{0.2cm} j \in N_{k}(i) \\
        0, & otherwise\\
        \end{cases},
\end{equation}
where $N_{k}(i)$ is the set of $k$ nearest neighbours of each log-return $Y_{t}(s_{i})$, and $\#N_{k}(i)$ is the cardinality of the set of $k$ neighbours of $i$. $W_{n}$ is then row-normalized by its construction. The number of order $p$ is chosen to be 1 for all time series; the model is estimated using $k = 3,5,7,9$. The estimated parameters and Bayesian Information Criterion (BIC) \citep{schwarz1978estimating} are shown in Table \ref{table:3}.

\subsection{Results and discussion}
The data involve 26 Asian stock indices and two US stock indices, namely NYSE and NASDAQ. Table \ref{table:3} shows the estimated parameters from fitting Markov-switching spatio-temporal ARCH models and one-regime spatio-temporal ARCH models with $k=3,5,7,9$ to the log-returns of 28 stock indices. Table \ref{table:3} shows that the models estimated with $k=5$ result in the lowest $BIC$ of 135053. Therefore, the parameters estimated using $k=5$ will be used for the rest of this case study. 

\begin{figure}[!ht]
    \centering
    \includegraphics[width=\textwidth]{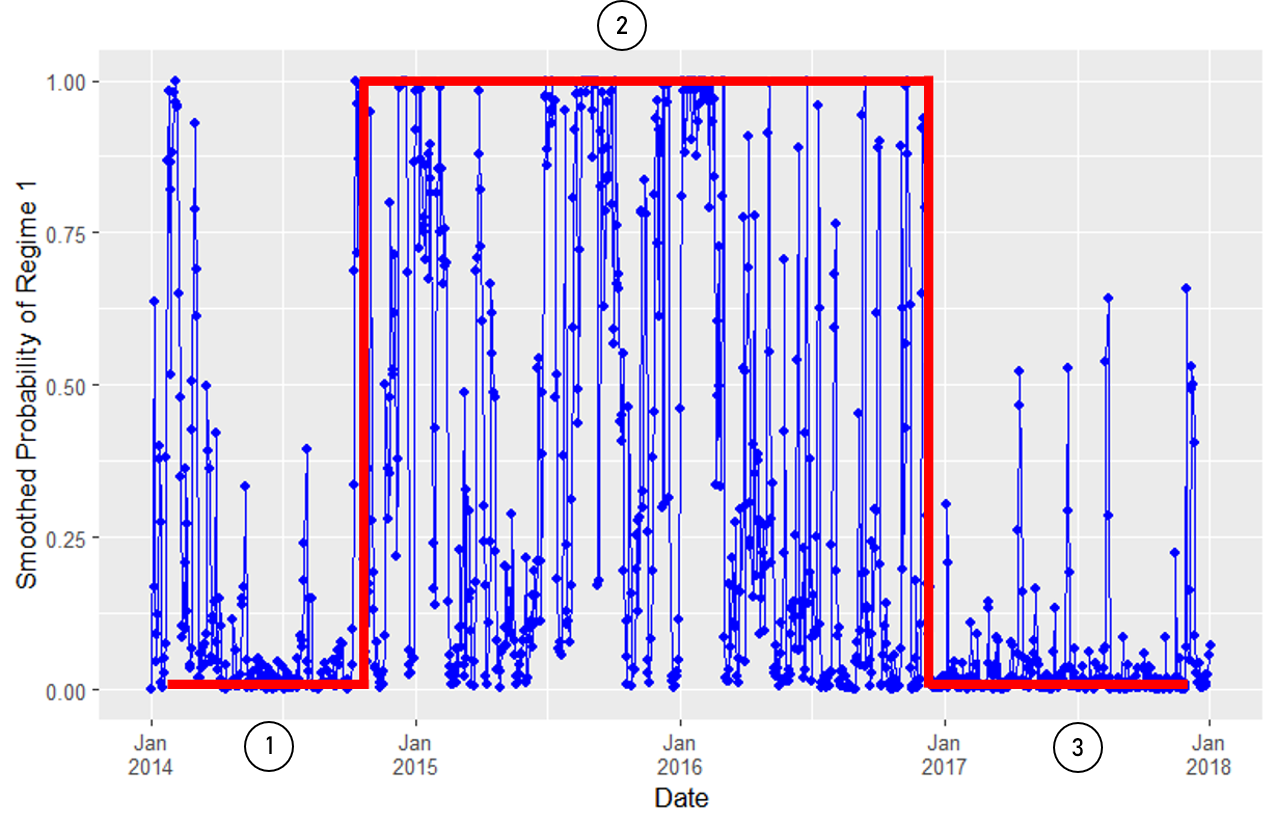}
    \caption{\small The blue curve represents the calculated smooth probability of Regime 1 (k=5); the red lines outline time periods which can be approximately categorized into 2 regimes based on the calculated smooth probability in regime 1. \textcircled{1} - The 2008-2014 quantitative easing program; \textcircled{2} - The Chinese stock market sell-off; \textcircled{3}  - Last regime-switching in the selected period.}
    \label{fig:smooth probs case study1}
\end{figure}

From Table \ref{table:3}, the high transition probabilities $p$ and $q$ suggest the existence of two regimes during this time period. The $BIC$ of the two-regime model shows improvement compared to the one-regime model, which further supports the existence of two regimes during the investigated time period. The two regimes can be differentiated primarily through the spatial autoregressive effects $\rho_{0,1}$ and $\rho_{0,2}$. The estimated values of  $(\rho_{0,1},\rho_{0,2}) = (0.260,0.073)$ suggest that the regime 1 can be viewed as a regime with more significant spatial effect. The temporal effects $\gamma_{0,1}$ and $\gamma_{0,2}$ are estimated to be positive and statistically significant. However, the spatio-temporal effect $\delta_{0,1}$ and $\delta_{0,2}$ are not statistically significant. This implies that spatial and temporal effects do not depend on each other. Figure \ref{fig:smooth probs case study1} shows the calculated smooth probability in regime 1 calculated for each data point. It is further observed that the 28 stock indices switch regime around September 2014 with a high probability. This could be attributed to the end of the quantitative easing initiative in the US in October 2014. An investigation by \cite{li2020volatility} indicates that the 2008–2014 quantitative easing program in the US had a positive spatial spillover effect on emerging markets like India and Indonesia. From Figure \ref{fig:smooth probs case study1}, it can be seen that the stock indices are in regime 1 from February 2015 to June 2015 with a high probability. This is followed by a structural break in June 2015 which leads to a regime-switching. In January 2017, the frequent regime-switching is observed to cease, and the stock indices are in regime 2 with a high probability.

To reiterate, the occurrences of regime-switching with high smooth probabilities are closely related to the different global events that occurred in the selected time period.

\begin{sidewaystable}
\captionsetup{singlelinecheck = false, format= hang}
\caption{Estimated parameters for Markov-switching spatio-temporal ARCH model of 28 stock indices.}
\label{table:3}
\begin{tabular}{@{\extracolsep\fill} c d{2.6} d{2.6} d{2.6} d{2.6} d{2.6} d{2.6} d{2.6} d{2.6} @{\extracolsep\fill}}
\toprule
  \multirow{3}{*}{\textbf{Parameter}}  & \multicolumn{2}{c}{${k=3}$} & \multicolumn{2}{c}{${k=5}$}
   & \multicolumn{2}{c}{${k=7}$}  & \multicolumn{2}{c}{${k=9}$}\\
  \cmidrule(lr){2-3} \cmidrule(lr){4-5} \cmidrule(lr){6-7} \cmidrule(lr){8-9}
   & \mc{\parbox{2cm}{\centering \textbf{Two-regime Estimate}}} & \mc{\parbox{2cm}{\centering \textbf{One-regime Estimate}}} 
   & \mc{\parbox{2cm}{\centering \textbf{Two-regime Estimate}}} & \mc{\parbox{2cm}{\centering \textbf{One-regime Estimate}}}
   & \mc{\parbox{2cm}{\centering \textbf{Two-regime Estimate}}} & \mc{\parbox{2cm}{\centering \textbf{One-regime Estimate}}}
   & \mc{\parbox{2cm}{\centering \textbf{Two-regime Estimate}}} & \mc{\parbox{2cm}{\centering \textbf{One-regime Estimate}}}
   \\ \midrule 
  $\rho_{0,1}$    &0.103^{***}    &0.108^{***} 
                  &0.260^{***}    &0.185^{***} 
                  &0.295^{***}    &0.220^{***} 
                  &0.307^{***}    &0.233^{***} \\
  $\gamma_{0,1}$  &0.202^{***}    &0.213^{***}  
                  &0.228^{***}    &0.208^{***}  
                  &0.228^{***}    &0.211^{***}  
                  &0.234^{***}    &0.212^{***}  \\ 
  $\delta_{0,1}$  &0.139^{***}    &0.083^{***}  
                  &0.033          &0.091^{***}  
                  &-0.001         &0.067^{***}  
                  &-0.046         &0.065^{***}  \\    
  $\mu_{0,1}$     &-4.05^{***}    &-5.33^{***} 
                  &-3.53^{***}    &-4.42^{***} 
                  &-3.69^{***}    &-4.25^{***} 
                  &-3.92^{***}    &-4.12^{***}  \\   
  $p$             &0.368^{\boldsymbol{\cdot}}      &\mc{-}       
                  &0.779^{*}      &\mc{-}  
                  &0.855^{***}    &\mc{-}       
                  &0.876^{***}    &\mc{-}  \\  
  $\rho_{0,2}$    &0.027^{**}     &\mc{-}       
                  &0.073^{***}    &\mc{-}  
                  &0.080^{***}    &\mc{-}       
                  &0.079^{***}    &\mc{-}  \\  
  $\gamma_{0,2}$  &0.206^{***}    &\mc{-}       
                  &0.185^{***}    &\mc{-}  
                  &0.186^{***}    &\mc{-}       
                  &0.185^{***}    &\mc{-}  \\
  $\delta_{0,2}$  &0.049^{***}    &\mc{-}       
                  &0.040^{*}      &\mc{-}  
                  &-0.004         &\mc{-}       
                  &-0.008         &\mc{-}  \\    
  $\mu_{0,2}$     &-6.97^{***}    &\mc{-}        
                  &-6.77^{***}    &\mc{-} 
                  &-7.20^{***}    &\mc{-}       
                  &-7.26^{***}    &\mc{-} \\    
  $q$             &0.836^{***}    &\mc{-}       
                  &0.920^{***}    &\mc{-}  
                  &0.938^{***}    &\mc{-}       
                  &0.946^{***}    &\mc{-} \\ 
  $BIC$           &\mc{135117}    &\mc{135515}  
                  &\mc{\underline{\textbf{135053}}}    &\mc{135334}  
                  &\mc{135067}    &\mc{135354}  
                  &\mc{135095}    &\mc{135368}   \\  
\bottomrule
\end{tabular}
\begin{tablenotes}
\item[] "." p < 0.1 "*" p < 0.05, "**" p < 0.01, "***" p < 0.001.
\end{tablenotes}
\end{sidewaystable}

\section{Conclusions}\label{Section 5}
To the best of our current knowledge, the integration of spatio-temporal ARCH models into a Markov-switching framework has not been previously explored. This research endeavours to bridge these gaps by presenting a novel model and proposing a quasi-maximum likelihood estimation method for its implementation. Through a comprehensive Monte Carlo simulation, we evaluate the finite sample properties of the maximum likelihood estimation approach. Stock-market data that potentially exhibits structural breaks was used to demonstrate the practicality of this proposed model. The results of the application to real data showcased the ability of the proposed model to efficiently detect regimes in spatially autocorrelated data. Future research focusing on network analysis to model structural breaks in stocks from diverse sectors, without consideration for geographical distance, is of particular interest. Moreover, a Markov-switching spatiotemporal ARCH model with more than 2 regimes can also be developed in the future.

\bibliographystyle{plainnat}
\bibliography{library.bib}
\end{document}